\def\etal{{\rm et al. }}

\def\kms{{\rm km\, s^{-1}}}
\newcommand\aap{{\em A}\&{\em A}}
\newcommand\aaps{{\em A}\&{\em AS}}
\newcommand\aj{{\em AJ}}
\newcommand\apj{{\em ApJ}}

\newcommand\apjs{{\em ApJS}}

\newcommand\mn{{\em MNRAS}}

\newcommand\pasp{{\em PASP}}

\documentclass[usenatbib]{mn2e}

\usepackage[dvips]{graphicx}
\usepackage{amssymb}
\usepackage{xcolor}    
\usepackage[normalem]{ulem}
\begin{document}

\title{Occurrence of LINER galaxies within the galaxy group environment}

\author[Coldwell \etal]{Georgina V. Coldwell$^{1}$, Luis Pereyra$^{2}$,
Sol Alonso$^{1}$, Emilio Donoso$^{3}$ and \newauthor Fernanda Duplancic$^{1}$\\
$^{1}$ Departamento de Geof\'{i}sica y Astronom\'{i}a, CONICET, Facultad de Ciencias Exactas, F\'{i}sicas y Naturales, Universidad Nacional \\
de San Juan, Av. Ignacio de la Roza 590 (O), J5402DCS, Rivadavia, San Juan, Argentina\\
$^{2}$ Instituto de Astronom\'ia Te\'orica y Experimental, CONICET, Observatorio Astron\'omico, Universidad Nacional de C\'ordoba, \\
C\'ordoba, Argentina\\
$^{3}$ Instituto de Ciencias Astron\'omicas de la Tierra y el Espacio, CONICET, Universidad nacional de San Juan, CC49, 5400,\\
 San Juan, Argentina}

\date{\today}

\pagerange{\pageref{firstpage}--\pageref{lastpage}}

\maketitle

\label{firstpage}

\begin{abstract}
We study the properties of a sample of 3967 LINER galaxies selected from SDSS-DR7, 
respect to their proximity to galaxy groups.
The host galaxies of LINER have been analysed and compared with a well defined 
control sample of 3841 non-LINER galaxies matched in redshift, luminosity, colour, 
morphology, age and stellar mass content.
We find no difference between LINER and control galaxies in terms of colour and
age of stellar population as function of the virial mass and distance to the geometric centre 
of the group.
However, we find that LINER are more likely to populate low density environments 
in spite of their morphology, which is typical of high density regions 
such as rich galaxy clusters. For rich (poor) galaxy groups, the occurrence of LINER 
is $\sim$2 times lower (higher) than the occurrence of matched, non-LINER galaxies.
Moreover, LINER hosts do not seem to follow the expected morphology-density 
relation in groups of high virial mass.
The high frequency of LINERS in low density regions could be due to the combination of
a sufficiently ample gas reservoir to power the low 
ionization emission and/or enhanced galaxy interaction rates benefiting 
the gas flow toward their central regions.
\end{abstract}


\section{Introduction}

Many features present in galaxy spectra, usually characterized by the intensity of absorption or emission lines, 
shape of continuum, etc., can provide important information about their formation and evolution.
In particular, emission lines probe the gaseous and 
chemical components of galaxies. The low-ionization nuclear emission-line regions 
(LINER) were described by \citep{Heck80} as a class of extragalactic objects 
with optical spectra dominated by enhanced low-ionization OI($\lambda 6300$) and 
NII($\lambda 6548,6583$) lines. Thus, LINER were defined by intensity ratios of 
optical emission lines, namely: (1) I([O II] $\lambda$ 3727)/I([O III] $\lambda$ 5007) $\geq$ 1 where [O II]
$\lambda$ 3727 is  used to designate the [O II] $\lambda$ $\lambda$ 3726, 3729
doublet, and (2) I([O I] $\lambda$ 6300) /I([O III] $\lambda$ 5007) $\geq$ 1/3.

The nature of the ionization source that powers emission lines has not been  
determined. In the last three decades different scenarios for the LINER excitation 
mechanism have been proposed. The most feasible are: (1) ionization by shock-heated 
gas due to direct mechanical energy input of turbulent gas motions, jets or bubbles
\citep{Heck80,DopSut95}; (2) stellar photo-ionization by hot O stars \citep{FilTer92} 
or by old post-asymptotic giant branch (AGB) stars \citep{Bin94,Stas08}; and (3) 
photo-ionization by an active central black hole \citep{Groves04}, such as an 
Active Galactic Nuclei (AGN).

The first mechanism, i.e. gas heated by fast shock waves, is probably the less 
favoured scenario because the velocity dispersion of nuclear gas commonly falls 
below the value required to explain the observed level of spectral ionization 
\citep{Ho03}. However, there is further controversy related to the last two 
mechanisms proposed. Observations at radio \citep{nagar05} and X-ray wavelengths 
\citep{GM09a} provide strong support for an AGN as the origin of LINER emission.
LINER seem to populate the low luminosity end of the AGN distribution \citep{Kewley06},
where radiatively inefficient accretion flows and external obscuring matter 
\citep{GM09b,dudik09} may cause the optical extinction. Thus, LINER are less 
luminous than Seyfert galaxies and share similar spectral characteristics, with 
the notable exception that LINER show enhanced low-ionization OI($\lambda 6300$) 
and NII($\lambda 6548,6583$) lines \citep{Heck80}. Furthermore, there is new
evidence supporting the hypothesis of stellar photo-ionization, as some authors 
suggest that hot post-AGB stars and white dwarfs could provide enough 
ionization to explain the LINER emission \citep{Bin94,sodre99,Stas08}. These stellar 
ionization sources are located in spatially extended regions around the nucleus 
as the $H\alpha$ and $H\beta$ brightness profiles do not decrease with $r^{-2}$,
i.e. the radial dependency expected for a central ionization source such as an AGN 
\citep{sarzi08,YanB12}.

LINER seems to be a common phenomena and they have been found in about 30\%
of nearby galaxies \citep{Heck80,HoFilSar97}. This percentage increases for galaxies 
with early type morphology, reaching 50\% in elliptical passive galaxies 
\citep{Fil86,Goud94,Yan06,CapB11}. This early galaxy type is consistent with that 
residing in dense galaxy environments. The well known morphology-age-density relation
\citep{Dress80,DML01} shows that red and massive early-type galaxies are associated 
to rich galaxy structures preferably located in the cores of groups or clusters 
of galaxies. On the other hand, late-type and blue galaxies are typically found 
in the outer regions of galaxy groups or in the field.

Since galaxy evolution correlates with different environmental factors, the 
study of the neighbourhood of LINER could provide important clues about the 
nature of their low ionization emission. If LINER are dominated by star-forming 
HII regions we would expect them to follow the morphology-density segregation \citep{Dress80}.
Instead, several recent studies have found that AGN environment do not seem to 
follow this relation \citep{PB06,coldwell09,padilla10,coldwell14}. Then, if the 
presence of an AGN is the main cause of the LINER emission we could assume that 
LINER will follow a similar trend that, for example, Seyfert 2 galaxies.

In addition, it is not clear that LINER constitute a homogeneous class of objects.
Unlike galaxies dominated by star-forming HII regions or Seyferts, whose ionization 
sources are clearly identified as young massive stars and AGN, respectively, 
LINER can be produced by a wide array of ionization mechanisms. For example, HST
observations have found that both unresolved nuclear and extended H$\alpha$ emission
are present in the majority of nearby LINER \citep{masegosa}, and LINER-like
ratios have long been observed in extended emission line regions \citep{Heck89,sarzi08}. 
In addition, \cite{graves07} found that red-sequence galaxies with LINER emission 
are younger than their quiescent counterparts, suggesting a connection between the star 
formation history and the mechanism generating the low ionization emission. But, there 
is no certainty about which mechanism prevails in LINER galaxies.
So the true power source of LINER emission is still under debate and most observations 
do not have resolution high enough to detect if the this emission comes from extended 
or nuclear regions of the galaxies.

The aim of this paper is to study the relation between LINER objects and their
local density environment in order to shed light on the origin of their main 
emission mechanism. The layout of this paper is as follows: in Section 2 we 
briefly describe the data selection and LINER classification scheme; in Section 3 
we describe the properties of galaxies hosting LINER, the procedure used to 
construct the control sample and their correlation with galaxy group parameters.
In Section 4 we investigate the occurrence of LINER in galaxy group and its dependency
with group proximity. Finally, in Section 5 we discuss the results and draw our 
conclusions. Throughout this paper, we have assumed a $\Lambda$-dominated 
cosmology, with $\Omega_{m} = 0.3$, $\Omega_{\lambda} = 0.7$ and 
$H_0 = 100 \kms Mpc^{-1}$.

\begin{figure*}
\includegraphics[height=60mm,width=160mm]{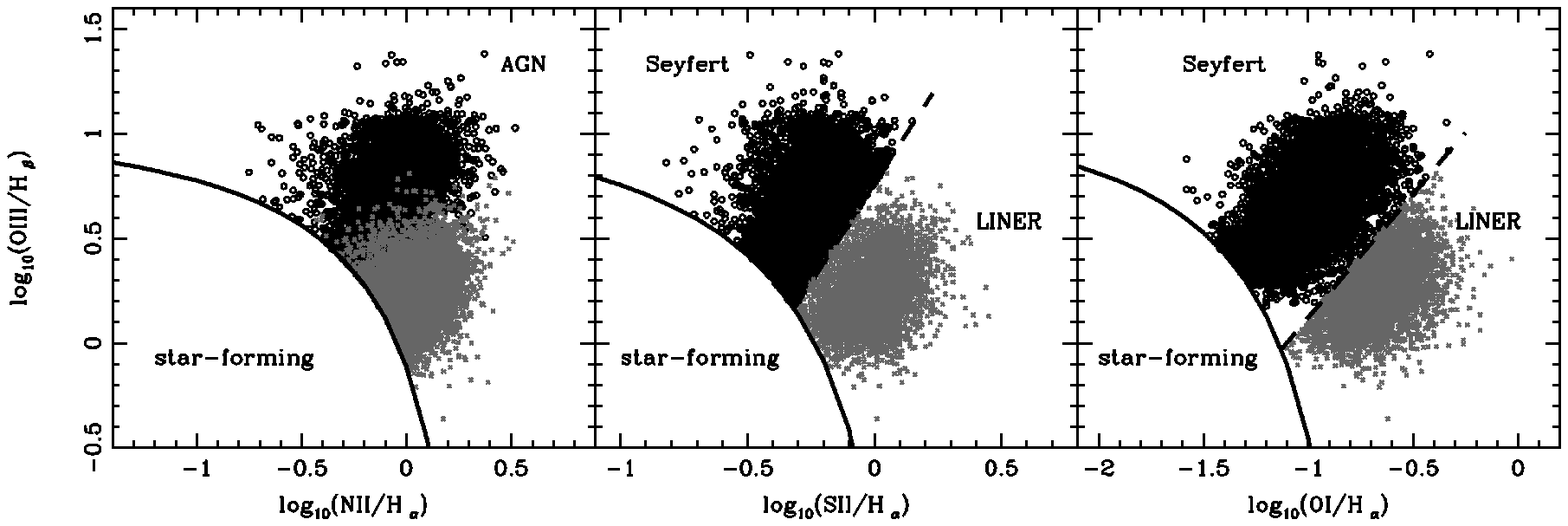}
\caption{BPT diagrams of the selection criteria defined by Kewley \etal (2006) 
used to classify emission-line galaxies as Seyfert or LINER.
The panels show the line ratios $\log([\rm OIII]/\rm H\beta)$ vs $\log(\rm [NII/H\alpha])$ (left),
$\log([\rm OIII]/\rm H\beta)$ vs $\log(\rm [SII/H\alpha]$ (central) and 
$\log([\rm OIII]/\rm H\beta)$ vs $\log(\rm [OI/H\alpha]$ (right).
Seyferts are indicated by black dots and LINER galaxies by grey crosses. The solid lines separate
star-forming galaxies from AGN and the dashed lines represent the Seyfert-LINER demarcation.}
\label{fig1}
\end{figure*}

\section{Data and Sample Selection}
In this paper we used galaxy samples selected from the spectroscopic Sloan 
Digital Sky Survey, Data Release 7 (SDSS-DR7, \cite{abazaja09}). The SDSS-DR7 
main galaxy sample is essentially a magnitude limited catalog \citep{petro76}
with \textit{$r_{lim}$}$ < 17.77$, consisting of $\sim 700000$ galaxies with 
measured spectra and photometry in five optical bands ({\it u,g,r,i,z}), and 
with a median redshift of $0.1$ \citep{strauss02}.

The procedures to estimate the physical properties of galaxies used in this analysis
are described by \cite{brinch04},\cite{tremonti04} and \cite{blanton05}. This data is
available from MPA/JHU\footnote{http://www.mpa-garching.mpg.de/SDSS/DR7/} and
NYU\footnote{http://sdss.physics.nyu.edu/vagc/}, including gas-phase 
metallicities, stellar masses, indicators of recent major star-bursts, current 
total and specific star-formation rates, emission-line fluxes and S\'ersic indices
among others.

\subsection{LINER Selection}
\label{sec:sel}

For the AGN selection we use the publicly available emission-line fluxes, whose
measurement is detailed in \cite{tremonti04}. Additionally, we have corrected 
for optical reddening using the Balmer decrement and the \cite{calzetti00} dust 
curve. We assume an $R_V=A_V/E(B-V)=3.1$ and an intrinsic Balmer decrement
$(H\alpha/H\beta)_{0}=3.1$ \citep{OM89}. Since the true uncertainties of the 
emission-line measurements were underestimated, the signal-to-noise ($S/N$) 
of every line was calculated with the flux errors adjusted as suggested by the MPA/JHU 
team\footnote{http://www.mpa-garching.mpg.de/SDSS/DR7/raw$\_$data.html}.

The emission-line galaxy sample was restricted to have a redshift range of 
$0.04 < z < 0.1$. The lower limit prevents that small fixed-size apertures 
affect galaxy properties derived from the fibre spectra, and the upper limit 
corresponds to the luminosity completeness limit for the SDSS sample. Furthermore, 
we only include galaxies with  $S/N > 2$ for all the lines involved in the three 
diagnostic diagrams used to discriminate Seyfert 2 from star-forming and LINER 
galaxies. This conservative criteria only moderately reduces the effective sample
size, yet assures a more reliable selection of LINER objects. Thus, from this 
refined sample, we separate Seyfert 2, LINER and star-forming galaxies using 
the three standard \cite[][: BPT]{BPT81} line-ratio diagrams. The AGN/starburst 
separation, as suggested by \cite{Kewley01,Kewley06}, depends on the relative 
source location within BPT diagnostic diagrams, and follows the equations

\begin{equation}
\log([\rm OIII]/\rm H\beta) > 0.61/(\log(\rm [NII/H\alpha])-0.47)+1.19,
\label{eqn:bpt1}
\end{equation}

\begin{equation}
\log([\rm OIII]/\rm H\beta) > 0.72/(\log(\rm [SII/H\alpha])-0.32)+1.30,
\label{eqn:bpt2}
\end{equation}

\begin{equation}
\log([\rm OIII]/\rm H\beta) > 0.73/(\log(\rm [OI/H\alpha])+0.59)+1.33,
\label{eqn:bpt3}
\end{equation}

LINER are located below the Seyfert/LINER division line as indicated by equations

\begin{equation}
\label{eqn:bpt4}
\log([\rm OIII]/\rm H\beta) < 1.89\log(\rm [SII/H\alpha])+0.76,
\end{equation}

\begin{equation}
\label{eqn:bpt5}
\log([\rm OIII]/\rm H\beta) < 1.18\log(\rm [OI/H\alpha]+1.30.
\end{equation}

On the grounds of internal consistency, we exclude from analysis ambiguous 
galaxies classified as one type of object in one BPT and another in the remaining
two diagrams. Bearing this in mind, we obtain an effective sample of 4778 LINER objects. 
The discriminated samples and the selection criteria are shown in the three BPT 
diagrams of Figure \ref{fig1}.

\subsection{Galaxy Group Catalogue}
In this work we use the group catalogue constructed by \cite{Zap09}, which were
identified by an improved version of the Huchra \& Geller (1982) friends-of-friends algorithm, 
with variable linking lengths of $D_{12}=D_0 R$ and $V_{12}=V_0 R$. These linking lengths are 
taken in the direction perpendicular and parallel to the line-of-sight, respectively. The spatial 
scaling $R$ takes into account the variation in the space density of galaxies in a 
flux-limited catalogue, and is calculated from the ratio of the density of galaxies brighter than 
the minimum required to enter the catalogue at the mean distance of the galaxies being linked, to
a characteristic survey depth. As found by \cite{MerZan05}, the algorithm parameters 
$D_0=0.239$h$^{-1}$Mpc and $V_0=450$kms$^{-1}$, are selected to produce a reasonably complete 
sample (95\%) with low contamination ($<$ 8\%).

In addition, the virial mass of galaxy groups are computed by means of the virial theorem, given by

\begin{equation}
\rm M_{vir} = \frac{\rm 3\sigma_v^2 \rm r_{vir}}{G},
\label{eqn:mvir}
\end{equation}

where $\sigma_v$ is the line-of-sight velocity dispersion and $r_{vir}$ is estimated as in \cite{MerZan05}, i.e.

\begin{equation}
\rm r_{vir} = \frac{\pi}{2}\frac{N_g(N_g-1)}{\sum_{i>j}r_{ij}^{-1}},
\label{eqn:rvir}
\end{equation}
 
where $N_g$ is the number of galaxy members and $r_{ij}$ are the relative projected distances 
between galaxies.

\section{Host galaxy properties}

The morphology of a galaxy can be indicative of evolutionary processes it has 
been subject to, and therefore an indicate of its environment. Spiral galaxies 
are known to be relatively unevolved galaxies residing typically in low 
density environments. In the other hand, elliptical galaxies are known to be
more evolved sources, as a result of dynamical processes such as mergers, and known to inhabit high density 
environments like galaxy groups or clusters \citep{toomre, dressler97}, where
they are subject to different effects such as strangulation, ram-pressure and 
close-encounters.

\begin{figure*}
\includegraphics[width=155mm,height=129mm]{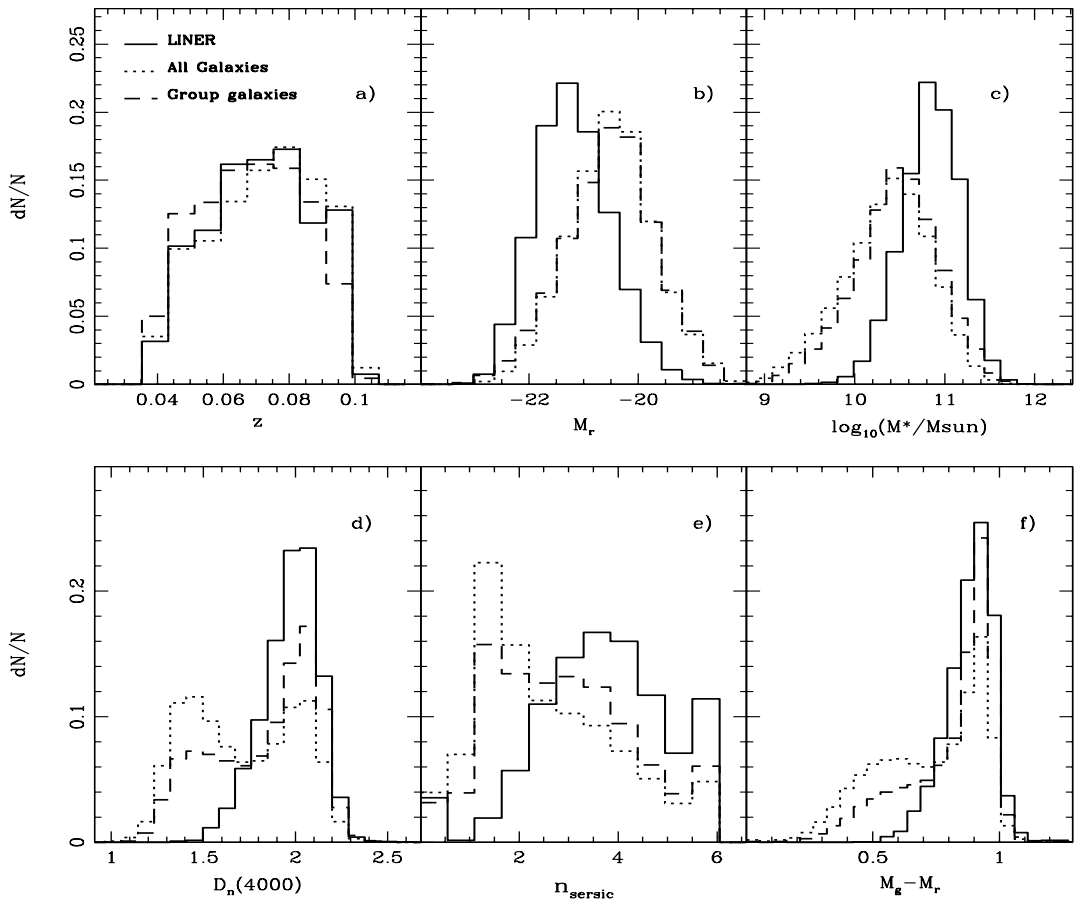}
\caption{Normalized distributions of galaxy properties for: SDSS-DR7 galaxies 
(dotted line), LINER total sample (solid line), and galaxy group members (dashed line).}
\label{dist}
\end{figure*}

The LINER host galaxies have particular characteristics which could give clues 
about their history. Figure \ref{dist} shows the normalized distributions of:
$a)$ Redshift, $b)$ (Petrosian) r-band absolute magnitude \citep{blanton03},
extinction and k-corrected bandshifted to $z=0.1$, using the software \texttt{k-correct\_v4.2} of \cite{Blanton2007}.
$c)$ stellar mass in logarithmic scale, $M^{\ast}$, 
previously determined by \cite{kauff03b} \footnote{The method relies on spectral
indicators relating to the stellar age and the fraction of stars formed in 
recent bursts.}, $d)$ $\rm D_n(4000)$ break index \citep{kauff02} $e)$ 
S\'ersic index \citep{sersic63}, and $f)$ colour ($M_g-M_r$).

The break index $\rm D_n(4000)$ is defined as the ratio of the average flux 
density in the narrow continuum bands ($3850-3950$ and $4000-4100$ $\AA$) and is 
suitably correlated to the mean age of the stellar population in a galaxy and can be 
used to estimate the star formation rate \citep{brinch04}. 
The majority of star formation takes place preferentially in galaxies with low
$\rm D_n(4000)$ values.

The change of surface brightness with distance from the galaxy centre can be well
described by the S\'ersic Law, which has a the form $\ln I(R)= \ln I_{0} -kR^{1/n}$.
Here, $I_{0}$ is the intensity at the centre and the S\'ersic index $n$ controls 
the degree of curvature of the profile. Setting $n = 4$ recovers a de Vaucouleurs profile, 
i.e. a good description of giant elliptical galaxies; while setting $n = 1$ recovers
the Freeman exponential profile, which is a good description of the light
distribution in disk galaxies and dwarf ellipticals. Most galaxies are well fit by S\'ersic 
indices in the range $0.5 < n < 10$.

Figure \ref{dist} also shows the distributions corresponding to the SDSS-DR7 main
galaxy sample (within the same redshift range). The bi-modality in the distribution of 
$\rm D_n(4000)$ and colour for the main galaxy sample is significantly noticeable. 
This effect is explained by the presence of two populations of galaxies in 
different stages of evolution: a population of young and blue galaxies, and another
composed of old and red galaxies. It is expected that the mean and variance of 
both colour distributions are dependent of the luminosity function and stellar mass 
\citep{Bernardi03,blanton03b,Hogg04} of the population considered. Moreover, 
Fig.\ref{dist} shows that most LINER are hosted by a single population of red 
and old galaxies. Consistently, these LINER hosts are significantly more massive 
and luminous than the general galaxy sample and have a bulge-type morphology.

These characteristics of LINER are in agreement with the results of \cite{Kewley06}
and \cite{CapB11}, and are consistent with very well known features of galaxy 
groups and cluster galaxy members \citep{Dress80, Dress84}.
We include in Figure \ref{dist} the distributions of SDSS-DR7 galaxy groups 
as constructed by \cite{Zap09}, noting that
galaxy group members show similar trends as the SDSS main sample, but with a 
higher percentage of red and old galaxies. This means that galaxy groups have 
a population of blue and young spiral galaxies, probably located in their outer 
regions, and another population of bright, red and bulge-type galaxies found 
in the centre of the systems. This is clear consequence of the morphological 
segregation \citep{DML01} of galaxies in clusters.

In this scenario, LINER galaxies can be thought as galaxies lying in the central 
area of rich galaxy groups and clusters. However, after cross-correlating the LINER 
catalogue with the catalogue of galaxy groups we found that 
only 12.4\% (592 objects) are members of galaxy groups 
with more than 10 members, within the redshift interval
considered. Instead of that, 20.4\% of LINER (974 objects) have been found to be a member of 
galaxy groups with a range of 4 to 9 members. This low percentage of LINER belonging to galaxy
groups richer than 10 members motivates us
to explore the relation between LINER properties and the group density environment.

\begin{figure*}
\includegraphics[width=175mm,height=135mm]{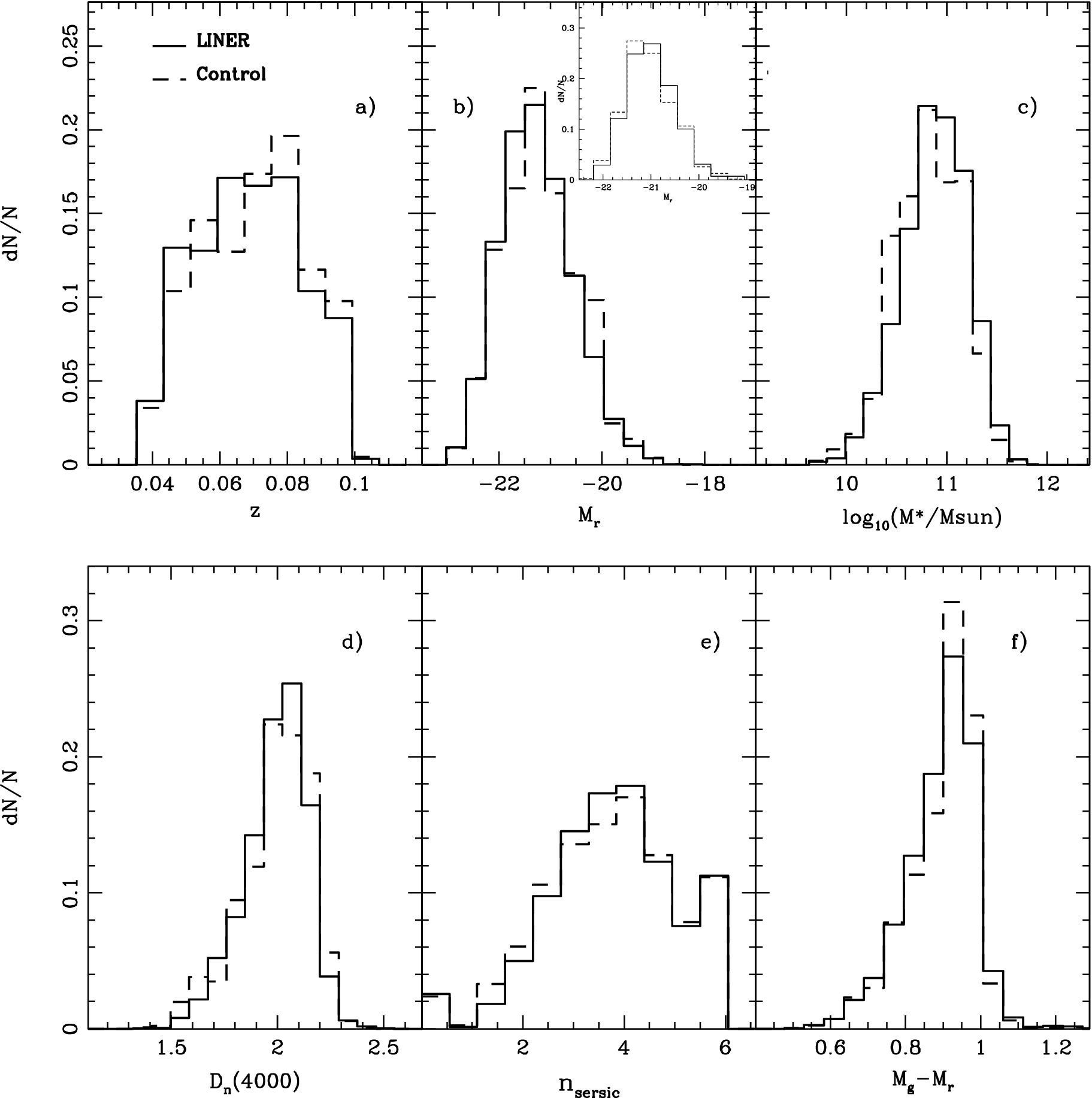}
\caption{Normalized distributions of galaxy properties for LINER (solid line) and control sample (dashed) belonging
to galaxy groups within $r_p < 2$ $r_{vir} $ and  $\Delta V < 1000$ $\kms$. The inner box in panel (b)
corresponds to the luminosity distribution of the brightest member of galaxy groups
matched in  both samples. 
}
\label{dist2}
\end{figure*}

\subsection{LINER close to Galaxy Groups and Construction of a control sample}
\label{sec:sel}
Galaxy groups are the most common structures in the universe and the connection 
between LINER and high density environments could provide valuable information 
about the mechanisms responsible of their low-ionization emission.

Considering the low occurrence rate of LINER in groups, we adopted as LINERS 
belonging to galaxy groups as those objects with projected distances $r_p < 2$ $r_{vir} $, 
where $r_{vir}$ is the virial radius, and radial velocity difference $\Delta V < 1000$ $\kms$, 
with respect to the geometric centre of the group. 
Thus, with this more relaxed criteria than only to be a galaxy group member  we are 
considering galaxies within the dark matter halo of the groups and, also, increasing 
the sample to obtain more confident statistical results.
In this way finally we obtained a sample of 3967 LINER galaxies (83\% of the total sample)
hosted by galaxy groups with more than 4 members.

To study the true environmental dependence of LINER properties it is very important
to select an appropriate control sample of galaxies without low-ionization emission
lines that belong to galaxy groups and fulfils the selection criteria as mentioned
before. The type of control samples have been used in a series of papers 
\cite{coldwell03,coldwell06} and \cite{coldwell09,coldwell14} to understand the 
behaviour of AGN with respect to non-active galaxies. In addition, by using SDSS 
mock galaxy catalogues built from the Millennium Simulation, \cite{Perez09} showed 
that a suitable control sample for galaxies in pairs should be selected (at least) 
with matched distributions of redshift, morphology, stellar mass, and local 
density environment. This criteria is also applicable to the case of control
galaxies for LINER. Thus, we construct our control sample by selecting
galaxies without low-ionization emission features from SDSS with matched distributions 
of those five parameters.

It is important to mention that the control sample has been taken from a galaxy
sample satisfying the criteria to be closer than $r_p < 2$ $r_{vir} $ and 
$\Delta V < 1000$ $\kms$ with respect to the galaxy group centre. Moreover, any 
galaxy classified as AGN or LINER has not been included in the control sample, 
which comprises 3841 galaxies. Figure \ref{dist2} shows the normalized 
distribution of properties for LINER host galaxies within the dark matter halos 
of galaxy groups, and their respective control sample. We can appreciate that 
the control sample suitably reproduces the properties of LINER objects. Both 
samples are redder, older and more massive than the rest of the galaxies in the 
SDSS catalogue, and have a S\'ersic index arithmetic mean of $~n = 4$.

\begin{figure}
\includegraphics[width=90mm,height=90mm,]{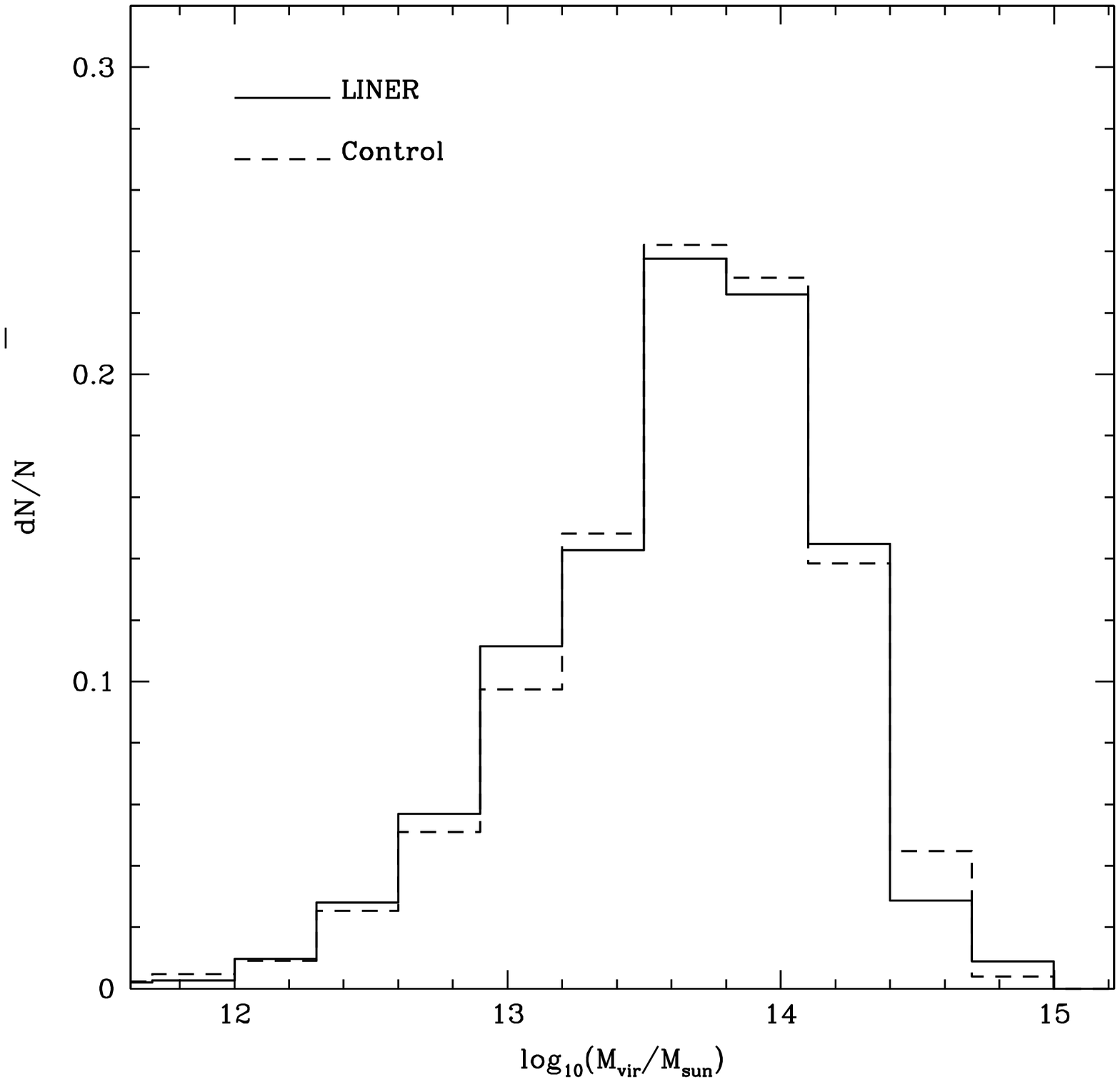}
\caption{Virial mass distributions for galaxy groups closer than $r_p < 2$ $r_{vir} $ and $\Delta V < 1000$ $\kms$
to LINER (solid line) and control sample (dashed line).}
\label{vmass3}
\end{figure}

In addition, we analysed the fraction of both LINER and control samples which correspond to the 
central galaxy of groups, identified as the brightest member galaxy. We found a similar percentage of central 
galaxies for control ($16.4\%$) and LINER ($13.8\%$) samples. The inner box in Figure 3(b) shows that the 
luminosity distribution of the central galaxies in both samples is very similar. Moreover, in Figure \ref{vmass3}
we also compare the virial mass distribution of groups close ($r_p< 2$ $r_{vir} $ and $\Delta V < 1000$ $\kms$) 
to galaxies of our LINER and control samples. The agreement between both distributions shows that, on average, 
LINER and control galaxies belong to galaxy groups of similar virial mass. As we will see in Section 4, this 
supports the idea that the difference in their fractions (Fig. 6 and 7) could be driven mainly by the 
presence of low-ionization emission in LINER hosts.

\subsection{Host galaxy properties: Dependency with the galaxy group proximity}

\begin{figure}
\includegraphics[width=90mm,height=90mm]{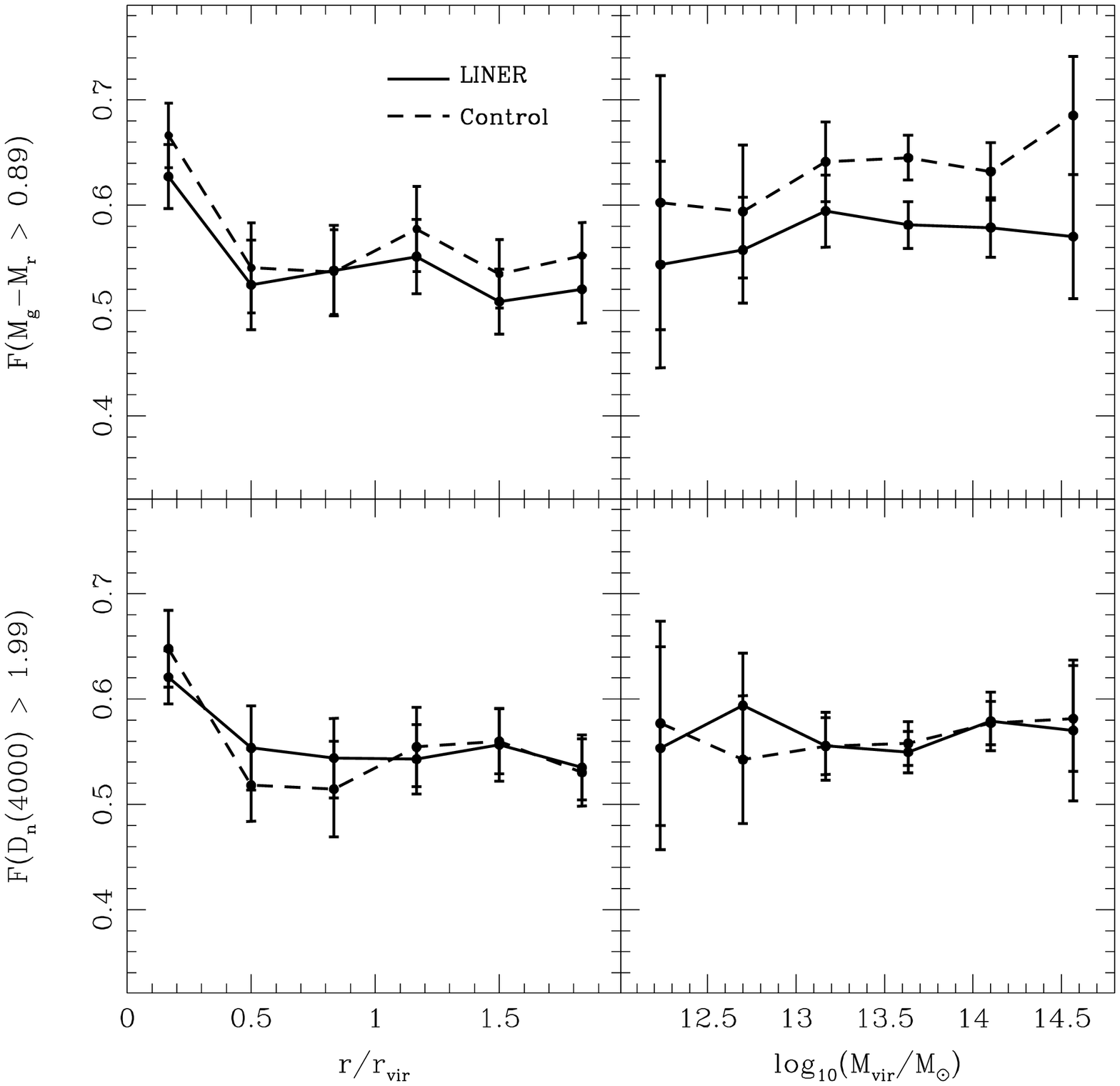}
\caption{Fraction of red ($M_g-M_r > 0.89$) and old ($D_n(4000)> 1.99$) galaxies
as function of the normalized projected distance to the group centre (left panels) 
and to the galaxy group virial mass (right panels). The solid line corresponds to
the LINER sample and the dashed line to the control sample.}
\label{lummass3}
\end{figure}
  
The similarity between LINER and control sample properties is what enables to explore 
how these are related with the galaxy group parameters with the aim to detect some 
distinctive feature for LINER. Particularly,  the galaxy colours provide an indirect 
constraint on the evolutionary history of galaxies since other galaxy parameters 
for example morphology, age, affect galaxy colours, resulting very appropriated for 
this study.

As in previous works \citep{coldwell06,coldwell09,coldwell14}, we use colours and 
the age indicator to study the fraction of host galaxies redder than 
$\rm M_g-M_r > 0.89$ and $D_n(4000) > 1.99$, as a function of the galaxy group 
centre distance and the virial mass. These thresholds for the fractions correspond 
to mean values of $\rm M_g-M_r$ and  $D_n(4000)$ for the LINER and control 
samples (see the distributions in Figure \ref{dist2}). The error bars are calculated with the bootstrap error 
resampling technique \citep{barrow84}. 

It is possible to observe a slightly higher fraction of red and old galaxies closer to 
the geometric group centre, as shown in Figure \ref{lummass3}. In addition, the trends 
of LINER and control samples are nearly indistinguishable.
 Moreover, other than a weak excess of red galaxies 
(still within the error bars), we do not find any significant difference with 
respect to the group virial mass. The conclusion is that the properties of LINER 
host galaxies at a given distance or virial mass do not vary respect to control
non-active galaxies, so the location within the galaxy group environment does not 
seem to affect colours and/or ages of LINER.

\section{Occurrence of LINER in galaxy groups}

In this section we explore the probability to find LINER respect to the control 
sample and the dependence with galaxy groups properties such as the virial mass, 
$M_{vir}$ (as derived from the virial theorem using the virial radius and 
velocity dispersion of group members). We also considered the group luminosity, 
$M_{r4}$, calculated by adding the r-band luminosities of the 4 brightest galaxy 
members in each group. This quantity, $M_{r4}$, is a better proxy for the true underlying 
group mass than the virial estimate \citep{Eke04,padilla04}. In this way, we
have considered two independent estimates of the group richness for this analysis. 

\begin{figure*}
\includegraphics[width=155mm,height=95mm,]{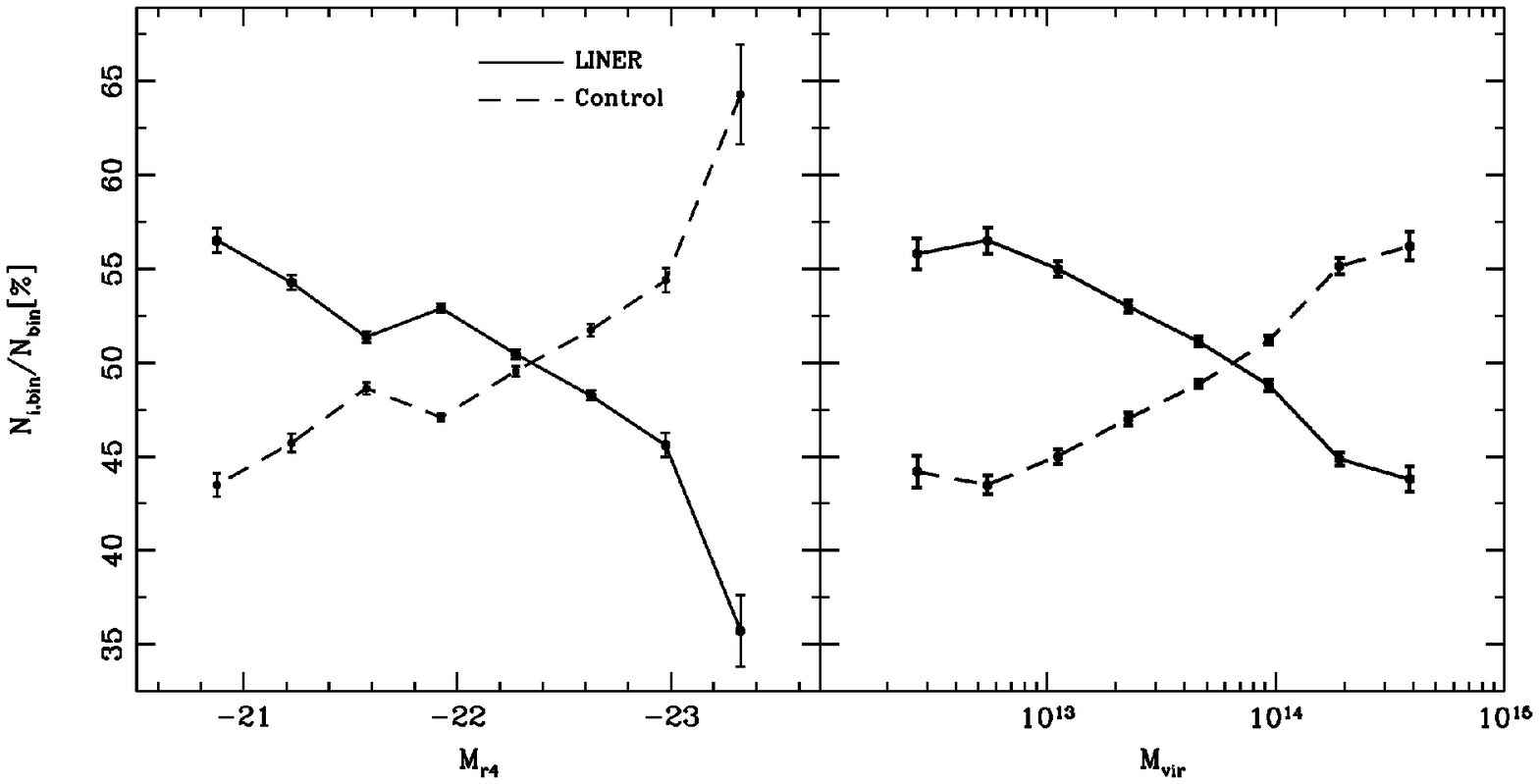}
\caption{Fraction of LINER (solid line) and the respective control sample (dashed lines) as a function
of the group virial mass $M_{vir}$ (right), and brightness $M_{r4}$ (left).}
\label{lummass}
\end{figure*}

Figure \ref{lummass} shows the fraction of both samples with respect to the 
total number of objects for a given bin of virial mass in logarithmic scale 
and group luminosity. We observe that the fraction of LINER 
drops with the group virial mass and luminosity, while the fraction of control 
galaxies increases with these parameters. Precisely for control sample, this is
expected due to the morphology-density relation and the sample distributions
shown in Figure \ref{dist2}. Yet somewhat surprisingly, the relation for LINER 
galaxies is in fact the opposite, even when their hosts galaxies are matched to the 
control sample. 
 
This tendency is more significant in the left panel of Figure \ref{lummass}, where 
the fraction of galaxies in the brightest groups in the control sample is 
approximately 2 times the LINER fraction. At low luminosities, the effect is 
is reversed as the proportion of LINER reaches its maximum value ($\sim$56\%)
and the difference respect to control falls to $\sim$13\%. 
A comparable trend is observed in the right panel of Fig. \ref{lummass}, with 
respect to $M_{vir}$, although the excess of control galaxies respect to LINER at 
higher values of $M_{vir}$ is somewhat lower than that shown for the group luminosity, 
$M_{r4}$.
These results indicate
that LINER galaxies have a preference to populate low mass, low luminosity 
environments formed by a higher proportion of blue galaxies with high gas content.

Bearing in mind the fact that several authors have found that the association 
between active galaxies and environment do not agree with the expected 
morphology-density relation \citep{PB06,coldwell09,padilla10,coldwell14}, 
these results suggest that the accretion of material inside the 
black hole should not be neglected as the main mechanism responsible of 
the low-ionization emission.

\begin{figure*}
\includegraphics[width=155mm,height=95mm,]{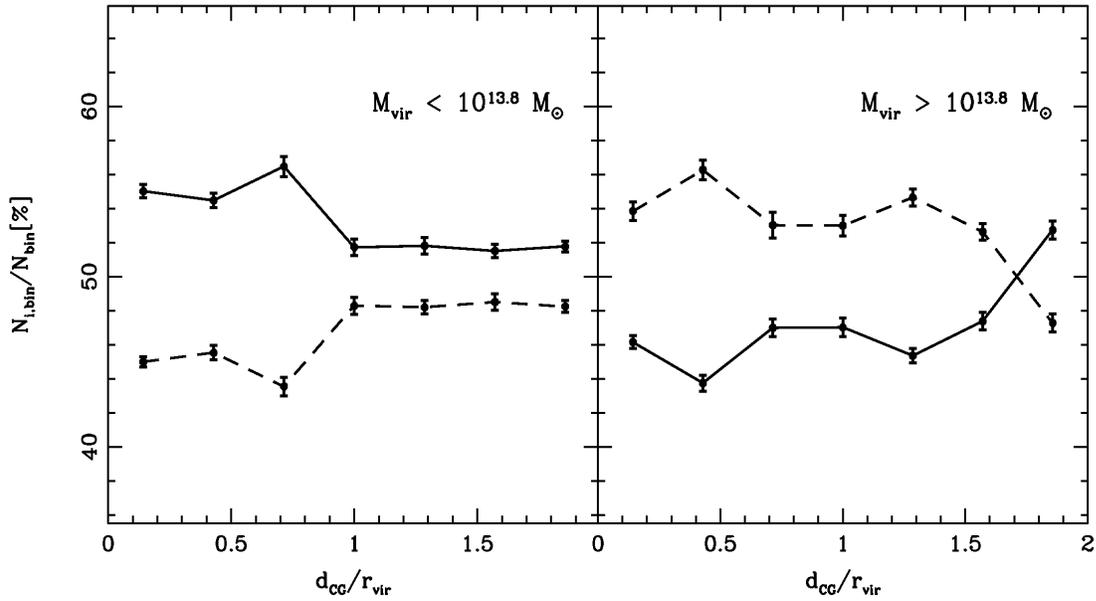}
\caption{Fraction of LINER (solid line) and the respective control sample (dashed lines) 
as a function of the normalized distance to the centre of galaxy groups with virial
mass $\rm log_{10}(M_{vir}) < 13.8$ (left) and $\rm log_{10}(M_{vir}) \geqslant 13.8$ (right).}
\label{lummass4}
\end{figure*}
  
Additionally, it is useful to perform this analysis by discriminating groups
with a group mass indicator. Figure \ref{lummass4} shows the fraction of LINER 
and control galaxies as a function of the distance to the galaxy group centre, 
normalized to the virial radius, $d_{cg}/r_{vir}$, for rich groups ($log_{10}M_{vir} > 13.8M_{\sun}$)
and for poor groups ($log_{10}M_{vir} < 13.8M_{\sun}$). In poor groups we observe a 
higher fraction of LINER respect to the control sample, which is not unexpected 
given Figure \ref{lummass}. This effect is almost independent of scale with only
a weak increase in the fraction of LINER galaxies toward the inner regions of 
poor galaxy groups. In rich groups, we find that the fraction of LINER increases
significantly toward the outer limits of these systems and that the fraction of
LINER objects is lower that in control galaxies.
Therefore, this reinforces the hypothesis suggested, that LINER galaxies prefer
to populate low density environments, where the probability of major mergers rises 
due to low galaxy velocity dispersions favouring this type of interactions.
This seems to be in agreement with previous results for AGN found in the literature
\citep{PB06,alonso07,coldwell09}.

\section{Discussion}

From SDSS-DR7 data we derived a large LINER catalogue at redshifts $0.04 < z < 0.1$.
As only a small fraction of LINER are hosted in groups of galaxies,
we adopted LINER objects with $r_p < 2 r_{vir}$ and $\Delta V < 1000 \kms$ as belonging 
to galaxy groups. In this way, our effective LINER sample comprises 3967 objects,
distributed in 2879 groups. The mean number of LINER per group
is $\sim 1.4$. In order to uncover the true environmental dependence 
of the characteristics of galaxies hosting LINER and get a hint
on the process responsible for generating low-ionization emission lines,
we constructed a control sample of non-active galaxies matched in redshift, 
luminosity, age, stellar mass, morphology and colour.
We also verified that : (1) the percentage of LINER and control galaxies corresponding to 
the central members of groups is quite similar between both samples; (2) galaxy groups nearby 
both samples have similar virial masses.

Following previous work, we have explored how host properties depend on mass and the proximity 
to galaxy groups. We have calculated the fraction of red ($M_g-M_r > 0.89$) and old 
($D_n(4000)> 1.99$) galaxies respect to the normalized projected distance, 
$d_{cg}/r_{vir}$, to the geometric group centre, and to the virial mass, $M_{vir}$.  
We find no significant difference in the tendencies of LINER and control samples. Therefore, 
the properties of hosts in both samples do not seem to vary respect to the location or mass of 
the galaxy groups.

In addition, we have carried out a detailed analysis the LINER occurrence within galaxy group 
halos and quantified the colour and age dependency with respect to representative 
galaxy group parameters such as radius, virial mass and group luminosity. We study 
the fraction of LINER with respect to two independent parameters representative 
of the galaxy group mass ($M_{vir}$ and $M_{r4}$). The results show a strong 
difference between LINER and control galaxies. While the latter fraction 
increases for rich groups, the fraction of LINER drops drastically. This 
effect is more significant for the highest luminosity groups, where we find about
2 times more control than LINER galaxies, indicating an evident preference of 
LINER to inhabit low mass galaxy groups.
Besides, LINER galaxies do not seem to follow the expected morphology-density 
relation in the proximity of rich galaxy groups.
Such a behaviour of LINER agrees with 
that corresponding to active objects (Seyfert 2, quasars, etc.) analysed in 
previous works. So, although we have no certainty about 
the degree at which the mechanisms are responsible of the low-ionization 
emission, the presence of nuclear activity is certainly
involved and should not be discarded.

In conclusion, we have shown the higher probability of LINER to be found in lower density 
environment, such as poor groups, where the galaxy interactions and content of gas are more abundant. 
Thus, the global conditions of these environments could favour the presence the LINER emission.

\section{Acknowledgments}
We would like to thanks to anonymous referee for the comments that helped to improve the paper.
This work was supported in part by the Consejo Nacional de 
Investigaciones Cient\'ificas y T\'ecnicas de la Rep\'ublica Argentina 
(CONICET), the Consejo Nacional de Investigaciones Cient\'ificas, T\'ecnicas y de Creaci\'on Art\'istica de la 
Universidad Nacional de San Juan (CICITCA) and the Secretaría de Estado de Ciencia, Tecnolog\'ia e Innovaci\'on 
del Gobierno de San Juan (SECITI).

Funding for the SDSS and SDSS-II has been provided by the Alfred P. Sloan 
Foundation, the Participating Institutions, the National Science Foundation, 
the U.S. Department of Energy, the National Aeronautics and Space Administration, 
the Japanese Monbukagakusho, the Max Planck Society, and the Higher Education 
Funding Council for England. The SDSS Web Site is \emph{http://www.sdss.org/}.
The SDSS is managed by the Astrophysical Research Consortium for the Participating 
Institutions. The Participating Institutions are the American Museum of Natural 
History, Astrophysical Institute Potsdam, University of Basel, University of 
Cambridge, Case Western Reserve University, University of Chicago, Drexel University,
Fermilab, the Institute for Advanced Study, the Japan Participation Group, Johns 
Hopkins University, the Joint Institute for Nuclear Astrophysics, the Kavli Institute 
for Particle Astrophysics and Cosmology, the Korean Scientist Group, the Chinese 
Academy of Sciences (LAMOST), Los Alamos National Laboratory, the Max-Planck-Institute 
for Astronomy (MPIA), the Max-Planck-Institute for Astrophysics (MPA), New Mexico 
State University, Ohio State University, University of Pittsburgh, University of 
Portsmouth, Princeton University, the United States Naval Observatory, and the University 
of Washington.

{}

\label{lastpage}

\end{document}